\documentclass[twocolumn,preprintnumbers,amsmath,amssymb,aps]{revtex4}

\usepackage[dvips]{color}
\usepackage{psfrag}
\usepackage{graphicx,graphics}
\usepackage{dcolumn}
\usepackage{bm}
\usepackage{amssymb}
\usepackage{slashed}
\usepackage{amsmath}
\usepackage{float}
\usepackage{pdfpages}
\graphicspath{{./Figures/}}

\def\lsim{\raise0.3ex\hbox{$\;<$\kern-0.75em\raise-1.1ex\hbox{$\sim\;$}}}
\def\gsim{\raise0.3ex\hbox{$\;>$\kern-0.75em\raise-1.1ex\hbox{$\sim\;$}}}

\newcommand{\be}{\begin{eqnarray}}
\newcommand{\ee}{\end{eqnarray}}

\newcommand{\n}{\nonumber\\}

\def\bea{\begin{eqnarray}}
\def\eea{\end{eqnarray}}
\usepackage{graphics}
\usepackage{array}
\begin{document}

\title{Higgs boson decays into  $\gamma\gamma$ and   $Z \gamma$ in the MSSM and BLSSM}
\author{A. Hammad$^{1}$, S. Khalil$^{1,2}$, S. Moretti$^{3}$ }
\affiliation{$^{1}$ Center for Fundamental Physics, Zewail City {of} Science and Technology,
6 October  City, Giza, Egypt.\\
$^{2}$ Department of Mathematics, Faculty of Science,  Ain Shams University, Cairo, Egypt.\\
$^{3}$ School of Physics \& Astronomy, University of Southampton, Highfield, Southampton, UK.}
\date{\today}

\begin{abstract}
We calculate Higgs decay rates into $ \gamma\gamma$ and $Z \gamma$  in the Minimal Supersymmetric Standard Model
(MSSM) and  (B-L) Supersymmetric
Standard Model (BLSSM) by allowing for contributions from light staus and charginos. We show that sizable departures are possible from the SM predictions for the 125 GeV state and that they are testable during run 2 at the Large Hadron
Collider. Furthermore, we illustrate how a second light scalar Higgs signal in either or both these decay modes can be accessed at the CERN machine rather promptly within the BLSSM, a possibility instead precluded to the MSSM owing to the much larger mass of its heavy scalar state.
\end{abstract}
\maketitle
The strongest experimental evidence of Higgs boson discovery at the Large Hadron
Collider (LHC) emerged from its decay channels into $\gamma \gamma$ and $ZZ$. Although these decays are at present
 largely consistent with the Standard Model (SM) predictions, one finds that the signal strength of the di-photon decay mode is larger than the SM expectation by an $\approx2 \sigma$ deviation \cite{Aad:2014eha,CMS:2014iua}. While this effect may well be compatible with the SM, the difference calls for close scrutiny, as a Higgs decay into di-photons is a loop-mediated process, thus subject to Beyond the SM (BSM) effects entering at the same perturbative level as the SM ones. Hence, it may well be
regarded as a possible hint of new physics.  In addition, both  ATLAS \cite{Aad:2014fia} and CMS \cite{Chatrchyan:2013vaa} reported upper bounds for the $Z \gamma$ decay rate which are
one order of magnitude larger than the SM expectation, thereby not eliminating
the possibility of deviations from the SM in this channel either. Indeed, just like $\gamma\gamma$, also $Z\gamma$ is induced by loops wherein BSM particles may enter alongside the SM ones.
Therefore, both such decay channels are key to understand the nature of the SM-like Higgs boson discovered at CERN
in July 2012 and they will be analysed very thoroughly in the second LHC run.

A common feature of the $\gamma\gamma$ and $Z\gamma$ decay modes is that they are both primarily mediated
by $W$-boson and $t$-quark loops, which are of opposite sign and with the former dominanting the latter, so that,
upon accounting for the dominance of the $h\to WW$ decay starting from $\approx160$ GeV, one finds that
the corresponding  Branching Ratios (BRs)
tend to be largest below the $WW$ threshold, say, around 130 and 150 GeV, respectively.
Another peculiarity of these two decay modes is that any contribution to the $\gamma \gamma$ channel
will affect the $Z \gamma$ one as well. The vice versa is not true though. For example, scenarios with a
$Z'$ boson which can mix with the $Z$ state of the SM would affect the latter but not the former. A spectacular
situation which would definitely hint at new physics is, for example, the one where the SM-like
Higgs decay rate into $Z \gamma$ is measured to be larger than the one in $ \gamma \gamma$. Recall in fact
 that for the Higgs boson of the SM with a 125 GeV mass one has that
$\Gamma(h \to \gamma \gamma)$ $>$ $\Gamma(h \to Z \gamma)$. Needless to say,
the discovery of another Higgs boson signal, in $\gamma\gamma$, $Z\gamma$ or else, would be a clear evidence
for a BSM nature of  Electro-Weak Symmetry Breaking (EWSB).

Amongst models of Supersymmetry (SUSY), a theory well placed as prime candidate for BSM physics, two are of interest
here. Firstly, the Minimal Supersymmetric Standard Model (MSSM), which contains two CP-even neutral Higgs bosons, $h$, the SM-like Higgs, and $H$, a much heavier state.  Secondly, the  (B-L) Supersymmetric Standard Model (BLSSM), which is an extension of the MSSM obtained via
enlarging its gauge group by a $U(1)_{B-L}$  and is one of the best motivated non-minimal SUSY models as it accounts for non-zero neutrino masses. The Higgs sector of the BLSSM consists of two Higgs doublets and two Higgs singlets ((B-L) charged). Therefore, one finds that the physical CP-even neutral Higgs bosons are four, $h, H, h'$ and $ H'$, where the first two are MSSM-like and the last two are the truly BLSSM ones. Of relevance in the choice of these two  benchmark
SUSY scenarios is the following, that, owing to the fact that they have the same quantum numbers and
$U(1)_Y$ and $U(1)_{B-L}$ are not orthogonal, the $Z$ boson of the SM and the $Z'$ of the BLSSM mix (and,
not less importantly, so do their SUSY counterparts),  a phenomenological aspect of course missing in the MSSM.
Furthermore, due to possible large mixing in the CP-even Higgs mass matrix, which is in turn proportional to the gauge coupling mixing between $U(1)_Y$ and $U(1)_{B-L}$, the Higgs boson $h'$ can become the second lightest Higgs
state  with mass $\gsim 135$ GeV \cite{Abdallah:2014fra}. Therefore, the BLSSM offers another
Higgs state which can  have significant decays into $\gamma\gamma$ and $Z \gamma$, in addition to $ZZ$, so that it may even explain a possible second Higgs peak at $\approx1$ GeV  in the CMS  samples of $\gamma\gamma$ and $ZZ$ \cite{Abdallah:2014fra}.

In previous analyses \cite{Belyaev:2013rza,Hemeda:2013hha,Casas:2013pta}, it was emphasised the role that  light stau and chargino effects can have onto the
di-photon decay rates in the MSSM. We revisit here those analyses by also including an investigation of
the $Z\gamma$ channel in the MSSM. Furthermore, we contrast these results with what instead emerges in the BLSSM. The aim is to assess whether significant differences may occur between the MSSM and BLSSM in the
$\gamma\gamma$ and/or $Z\gamma$ decay channels with respect to the SM and indeed between each other. Finally, we also intend to establish the LHC scope in accessing one or the other of
these two modes when the decaying object is the lightest genuinely BLSSM Higgs state, thereby ultimately enabling one
to claim a possible evidence of SUSY and, at the same time, to confirm its non-minimal form.

As intimated, just like for the case of the $h\to \gamma\gamma$ decay (whose formulae can be found in \cite{Belyaev:2013rza,Hemeda:2013hha,Djouadi:1996pb}),
in the MSSM. a significant effect onto the decay width of $h\to Z\gamma$ may be obtained through the exchange of a light stau and/or light chargino. For this mode, the partial decay  width is given by \cite{Lee:2007gn}:
\bea
\label{eq:2.1}
\Gamma(h\to Z\gamma) &=& \frac{G^2_F \alpha^2 M^2_W m^3_h}{64\pi^4}\left(1-\frac{M^2_Z}{m^2_h}\right)^3  \nonumber\\
&\times&\left|A_t + A_W + A_{\chi^\pm} + A_{\tilde{\tau}}\right|^2,~
\eea
where $G_F$ is the Fermi constant. The SM form factors $A_t$ and $A_W$ are obtained from the loops mediated by the  $t$-quark and $W$-boson, respectively. The explicit form of $A_{t,W}$ can be found in Ref. \cite{Djouadi:1996yq}. The SUSY form factor  $A_{\tilde{\tau}}$ is given by
\be
\label{eq:2.4}
A_{\tilde{\tau}} = \frac{4 \upsilon^2}{c_W s_W M_{\tilde{\tau}}^2}\sum_{ij} g_{h\tilde{\tau}_i\tilde{\tau}_j} g_{Z\tilde{\tau}_i\tilde{\tau}_j} ~C_2(M_{\tilde{\tau}_i},M_{\tilde{\tau}_j},M_{\tilde{\tau}_j}) ,
\ee
where $g_{Z\tilde{\tau}_i\tilde{\tau}_j}$ and $g_{h\tilde{\tau}_i\tilde{\tau}_j}$ are the couplings of the $Z$ and
$h$ boson to staus, respectively. Now, the stau mass matrix can have a large mixing if $A_\tau$ or $\mu
\tan \beta$ is large enough, Therefore, one of the eigenvalues, say,
$M_{\tilde \tau_1}$, can be as light as $100$ GeV.
The Higgs coupling to the lightest stau, normalised by
$v/\sqrt{2}$, with $v$ the SM Higgs Vacuum Expectation Value (VEV),
is%
\bea%
g_{h\tilde{\tau}_1\tilde{\tau}_1} &=& -\frac{1}{2} \cos^2
\theta_{\tilde{\tau}} +  \sin^2 \theta_W \cos 2
\theta_{\tilde{\tau}}  - \frac{M_{\tau}^2}{M_Z^2} \nonumber\\
&-&\frac{M_{\tau}(A_\tau - \mu \tan \beta)}{2 M_Z^2} \sin
2 \theta_{\tilde{\tau}}.%
\eea %
With a large stau mixing, $\sin 2 \theta_{\tilde{\tau}} \simeq 1$,
 $\tan \beta > 50$ and $\mu \sim$ TeV one finds that %
$ g_{h\tilde{\tau}_1\tilde{\tau}_1} \simeq \frac{M_{\tau}
\mu \tan \beta}{2 M_Z^2}.$ %
Therefore, the sign of the stau contribution depends on the sign
of $\mu$. Finally the loop function $C_2(M_{\tilde{\tau}_i},M_{\tilde{\tau}_j},M_{\tilde{\tau}_j})$ is again given in Ref. \cite{Djouadi:1996yq}.

\begin{figure}[t]
\includegraphics[height=6cm,width=7.75cm]{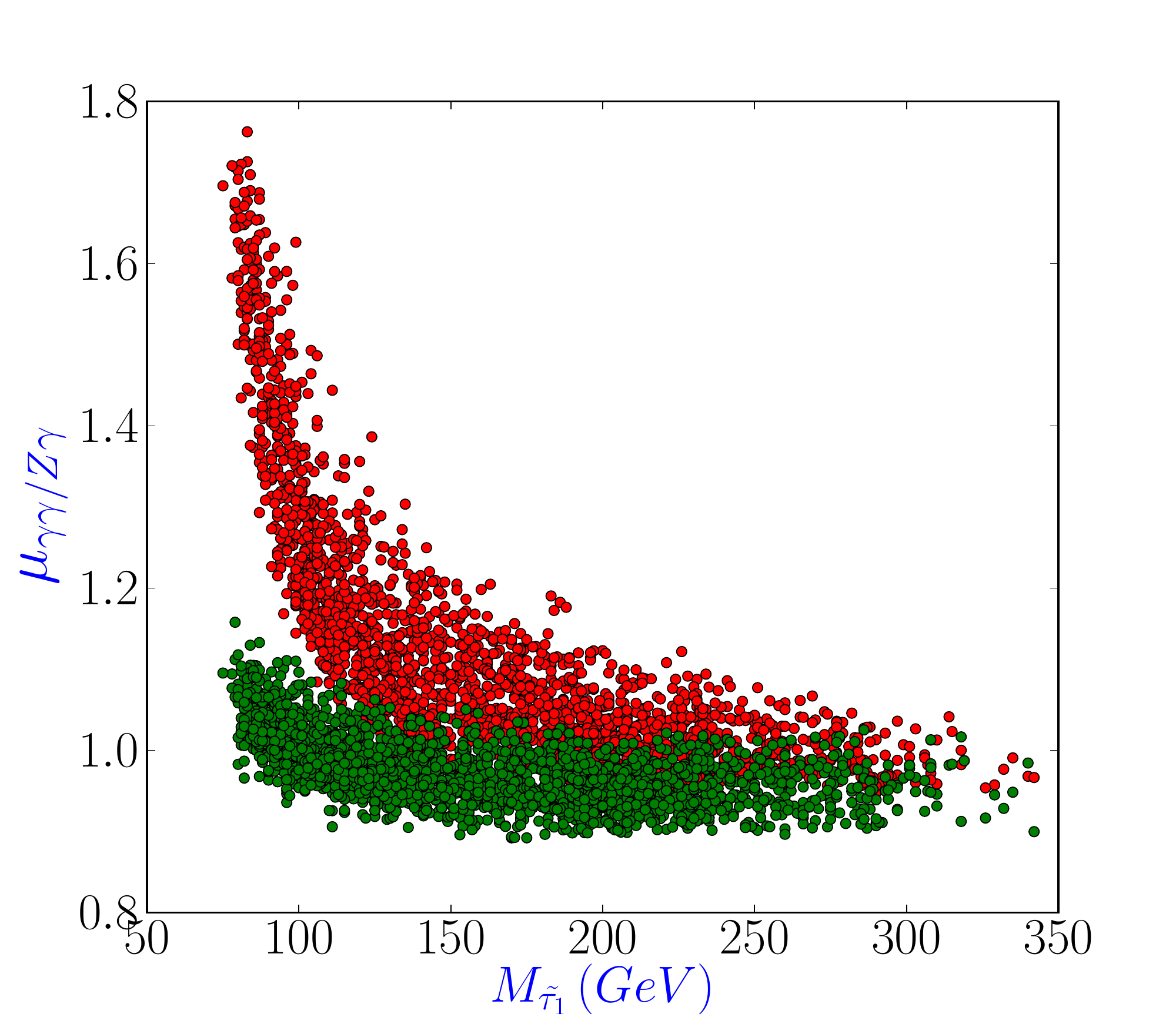} ~ \includegraphics[height=6cm,width=7.75cm]{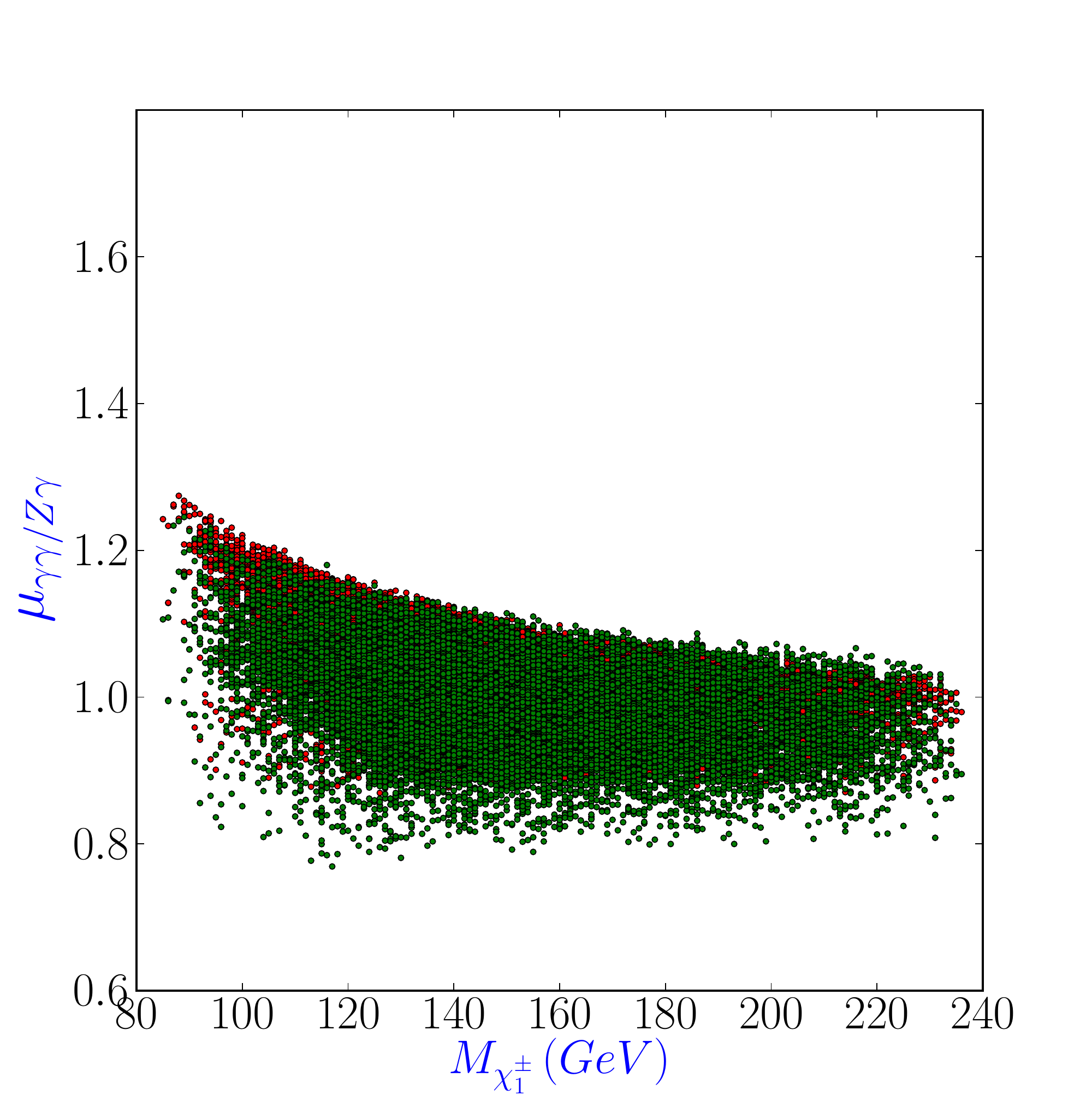}
\caption{Signal strength of $h\to \gamma\gamma$  (red) and $Z\gamma$ (green) versus the lightest stau (top) and
 chargino (bottom) mass.  }
\label{fig:1}
\end{figure}

As mentioned, also the charginos can mediate $h\to Z \gamma$ and they too  can be light, ${\cal O}(100)$ GeV. The chargino form factor $A_{\chi^\pm_{ij}}$ is given by
\bea
\label{eq:2.5}
A_{\chi^\pm_{ij}} &=& - 2\sqrt{2}M^2_Z \cot_W \sum_{ij} \frac{M_W}{M_{\chi^\pm_{ij}}}  g_{Z\chi^+_i \chi^-_j} g_{h\chi^+_i \chi^-_j}
\nonumber \\
&\times& f(M_{\chi^\pm_{i}},M_{\chi^\pm_{j}},M_{\chi^\pm_{ij}}),\ee
where $g_{Z\chi^+_i \chi^-_j}$ and  $g_{h\chi^+_i \chi^-_j}$ are the couplings of the $Z$ and $h$  to charginos,
respectively.
Note that, due to the vector and axial interactions of the $Z$ boson, both diagonal and off-diagonal couplings of charginos can contribute to the $h\to Z\gamma$ decay. The couplings of the Higgs boson $h$ with charginos are given by $g_{h \chi_i^+\chi_j^-}  =C^L_{ij} P_L + C^R_{ij} P_R $, where
\bea%
C^L_{ij} &=& \frac{1}{\sqrt{2}sw}\left[-\sin \alpha V_{j1} U_{i2} + \cos \alpha V_{j2} U_{i1} \right],\\
C^R_{ij} &=& \frac{1}{\sqrt{2}sw}\left[-\sin \alpha V_{i1} U_{j2} + \cos \alpha V_{i2} U_{j1} \right].%
\eea%
These couplings can reach their maximum values and become of order ${\cal O}(\pm 1)$ if $\tan \beta$ is very small, close to one, and $\mu \simeq M_2$. In Ref. \cite{Hemeda:2013hha}, it was emphasised that the Higgs couplings to charginos can be negative, hence the chargino can give a constructive interference with the $W$-boson  that may lead to a possible enhancement for $\gamma\gamma$. In $ Z \gamma$ too  the relative sign of $g_{Z\tilde{\chi_i}^+\tilde{\chi_j}^-}$ and $g_{h\tilde{\chi_i}^+\tilde{\chi_j}^-}$ is important for enhancing (or suppressing)
the effective signal strength of the $hZ\gamma$ coupling. Finally, the loop functions  $f(x_1,x_2,x_3)$ can be found in \cite{Djouadi:2005gj}.

The signal strength of  $h  \to Z \gamma$, relative to the SM expectation, in terms of production cross section ($\sigma$) and decay BR, is defined as
\bea%
\!\!\!\mu_{Z\gamma}\!\!\!&\!\!\!\!=\!\!\!&\!\!\!\frac{\sigma(pp\!\to\! h\to\! Z\gamma)}{\sigma(pp \!\to\! h \!\to\!
Z\gamma)^{\rm{SM}}}\!\!=\!\!\frac{\sigma(pp \!\to\! h)}{\sigma(pp \!\to\!
h)^{\rm{SM}}}\frac{{\rm BR}(h\to Z\gamma)}{{\rm BR}(h \to Z\gamma)^{\rm{SM}}} \nonumber\\
&=& \frac{\Gamma(h \to gg)}{\Gamma(h \to gg)^{\rm{SM}}}
~\frac{\Gamma_{\rm{tot}}^{\rm{SM}}}{\Gamma_{\rm{tot}}}\frac{\Gamma(h \to Z
\gamma)}{\Gamma(h \to Z\gamma)^{\rm{SM}}} .%
\eea%
(A similar expression holds for $\gamma\gamma$.)
{In computing $\mu_{\gamma\gamma}$ and $\mu_{Z\gamma}$ we have used SARAH \cite{Staub:2008uz,Goodsell:2014bna} and SPheno \cite{Porod:2003um,Porod:2011nf} to build the model.  Then we linked it with CPsuperH \cite{Lee:2007gn,Lee:2012wa}
to compute the numerical values of the Higgs decays in all channels.}

In Fig. \ref{fig:1} we display the results of the signal strengths of  $h\to\gamma\gamma$  and  $Z \gamma$ as a function of the lightest stau and chargino masses for $m_h \simeq 125$ GeV. For the chargino plot, we scan over the following parameter space: $1.1 \leq \tan \beta \leq 5$, $100$~ GeV $< \mu < 300$~ GeV and $100$~ GeV $< M_2 < 300$~GeV.  For the stau plot, we randomise over the following parameter ranges: $5 \leq \tan \beta \leq 50$, $250 {\rm GeV} \leq m_{L_3,E_3} \leq 500$ GeV,  $500$~ GeV $< \mu < 2000$~ GeV, and $M_1=M_2=M_3=3000$ GeV. Other dimensionful SUSY parameters  are fixed to be of order few TeV so that all other possible SUSY effects onto $\mu_{\gamma \gamma}$ and $\mu_{Z\gamma}$  are essentially negligible.  As can be seen from the plots, the stau contribution may lead to a limited enhancement for the signal strength of $ \mu_{Z \gamma}$, 1.1 or so, unlike the case of  $\mu_{\gamma \gamma}$, which can be increased up to $1.6$ at $M_{\tilde{\tau}} \sim 100$  GeV. Curiously, it can happen, for large stau masses, that $\Gamma(h \to \gamma \gamma)$ $<$ $\Gamma(h \to Z \gamma)$.  The charginos instead contribute to   $ \mu_{\gamma \gamma}$ and $\mu_{Z \gamma}$ equally and both modes can be enhanced up to $1.2$.

\begin{figure}[t]
\centering
\includegraphics[width =0.5\textwidth]{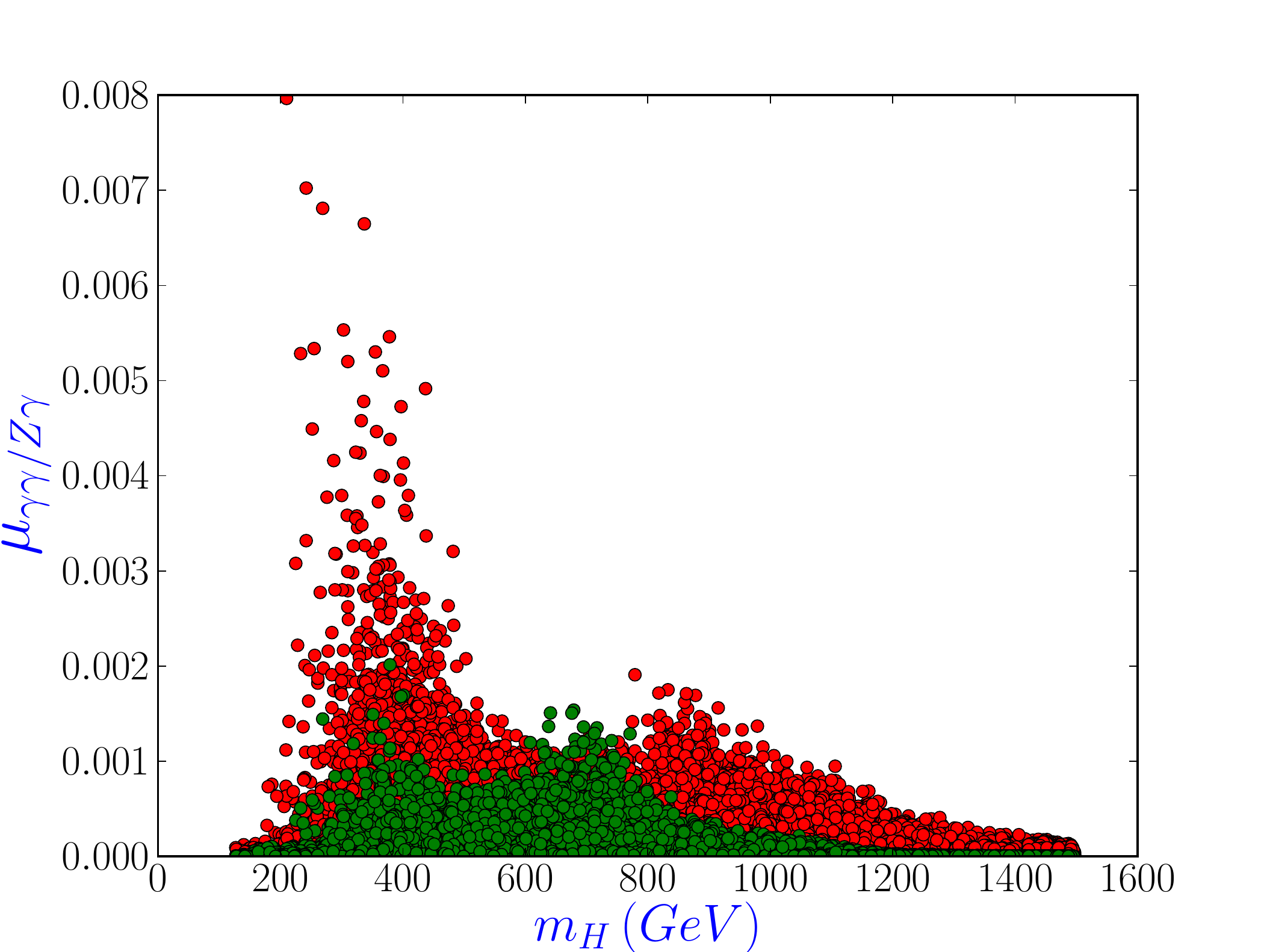} ~ \includegraphics[height=6cm,width=7.75cm]{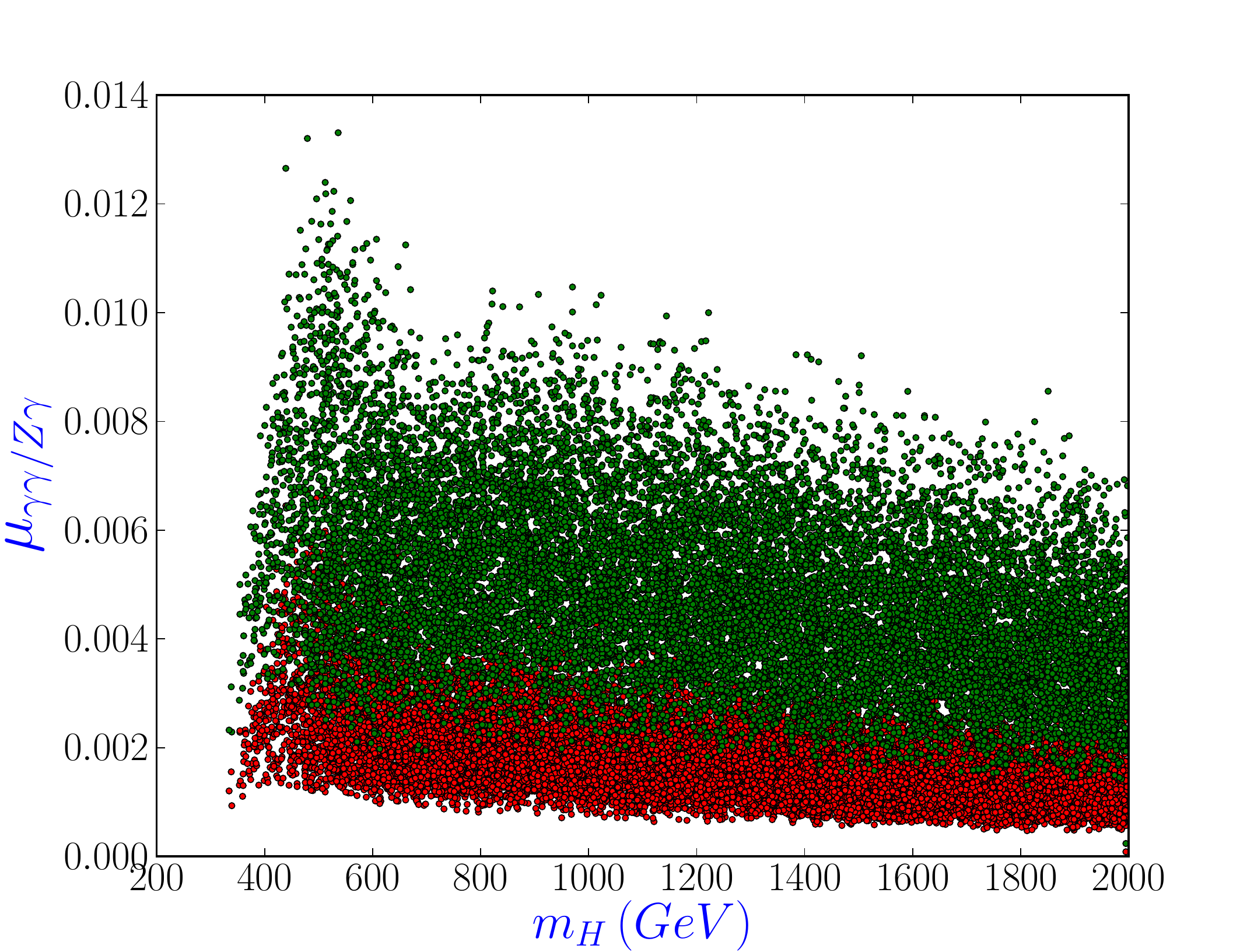}
\caption{\label{fig:2} Signal strength of $H\to \gamma\gamma$  (red) and $Z\gamma$ (green) versus the
$H$ mass for the
 light stau (top) and
light chargino (bottom) scenario. }
\end{figure}
Now we consider the decay of the MSSM heavy Higgs, which has a mass of order $m_H \sim (m^2_A +\sin^2 2\beta M^2_Z)^{1/2}$ and coupling with $W$ gauge-bosons equal to $g_{HWW} = - 2M^2_Z/M^2_A \tan^2 \beta$. It is clear that, for large $\tan\beta$ and  $m_A \gg  M_Z$  (as required for compliance with LHC data),  the coupling $g_{HWW}$ will have to be very small. Therefore, the main contribution to $H \to \gamma\gamma$ and $Z \gamma$ through $W$ exchange is significantly suppressed and hence one expects the corresponding decay rates
to be  much smaller than those of the SM-like Higgs, finally recalling the relative dominance of $H\to WW$ (as
$m_H>2 M_W$). This conclusion is confirmed by Fig. \ref{fig:2}, where we display the signal strength (again normalised
to the SM rates for $m_h=125$ GeV) of $H \to \gamma\gamma$ and $Z \gamma$ as a function of $m_H$.
Altogether, the signal strengths of $H \to Z \gamma$ and $H \to \gamma  \gamma$ are much smaller than 1, so probing these channels will be rather difficult. However, it is remarkable that  the signal strength of $H \to Z \gamma$ is larger than the
$H \to \gamma  \gamma$ one throughout the entire $H$ mass interval considered, though the phenomenological relevance of this is dubious, given the poor event rates overall. We trace this effect back as being due to the light stau contribution,   associated to very large values of $\tan \beta$.
In contrast, for a light chargino yielding an enhancement occurring for $\tan\beta < 5$, $\gamma\gamma$ is
generally more sizable that $Z\gamma$.

\begin{figure}[t]
\includegraphics[height=6cm,width=7.75cm]{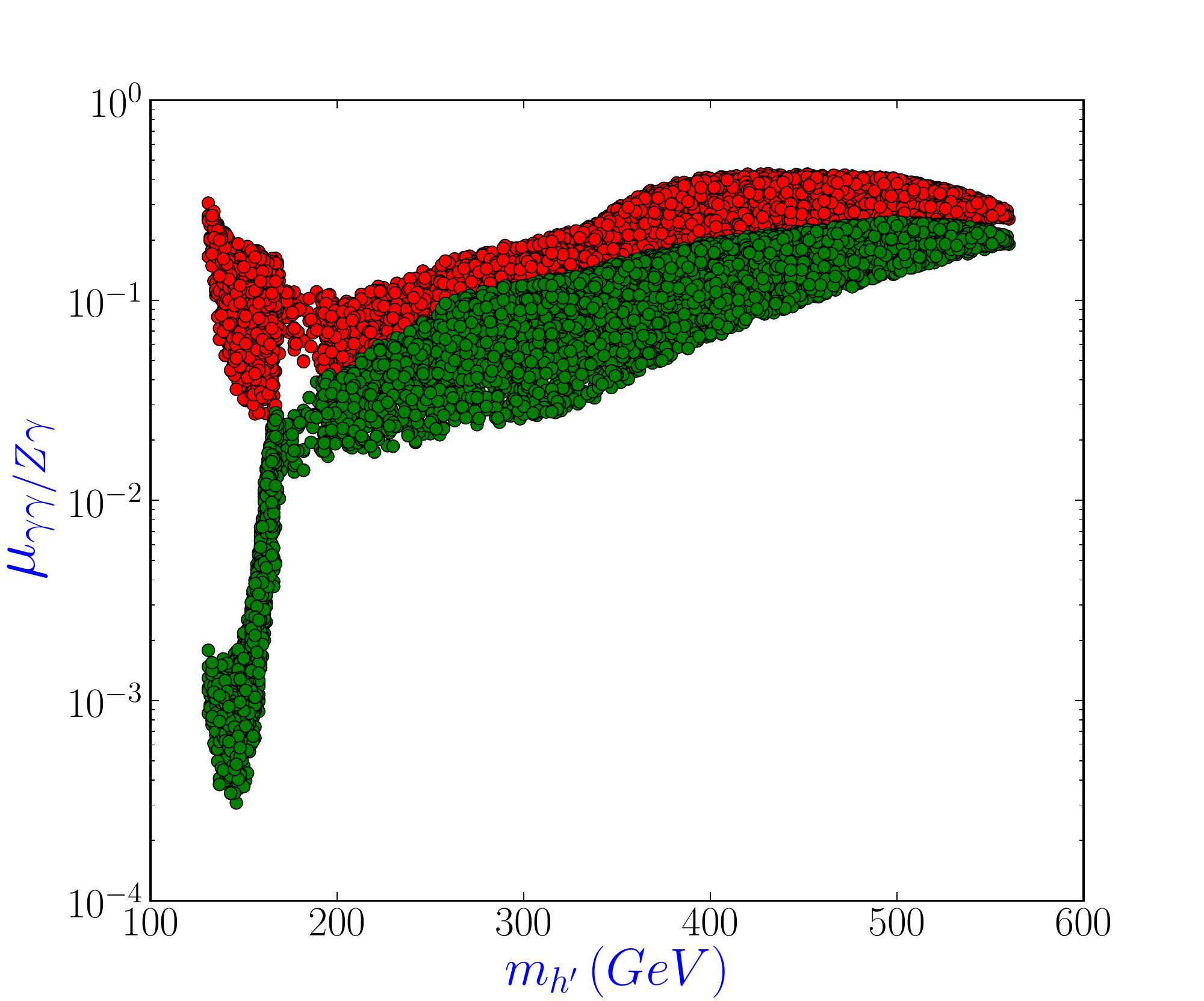} ~ \includegraphics[height=6cm,width=7.75cm]{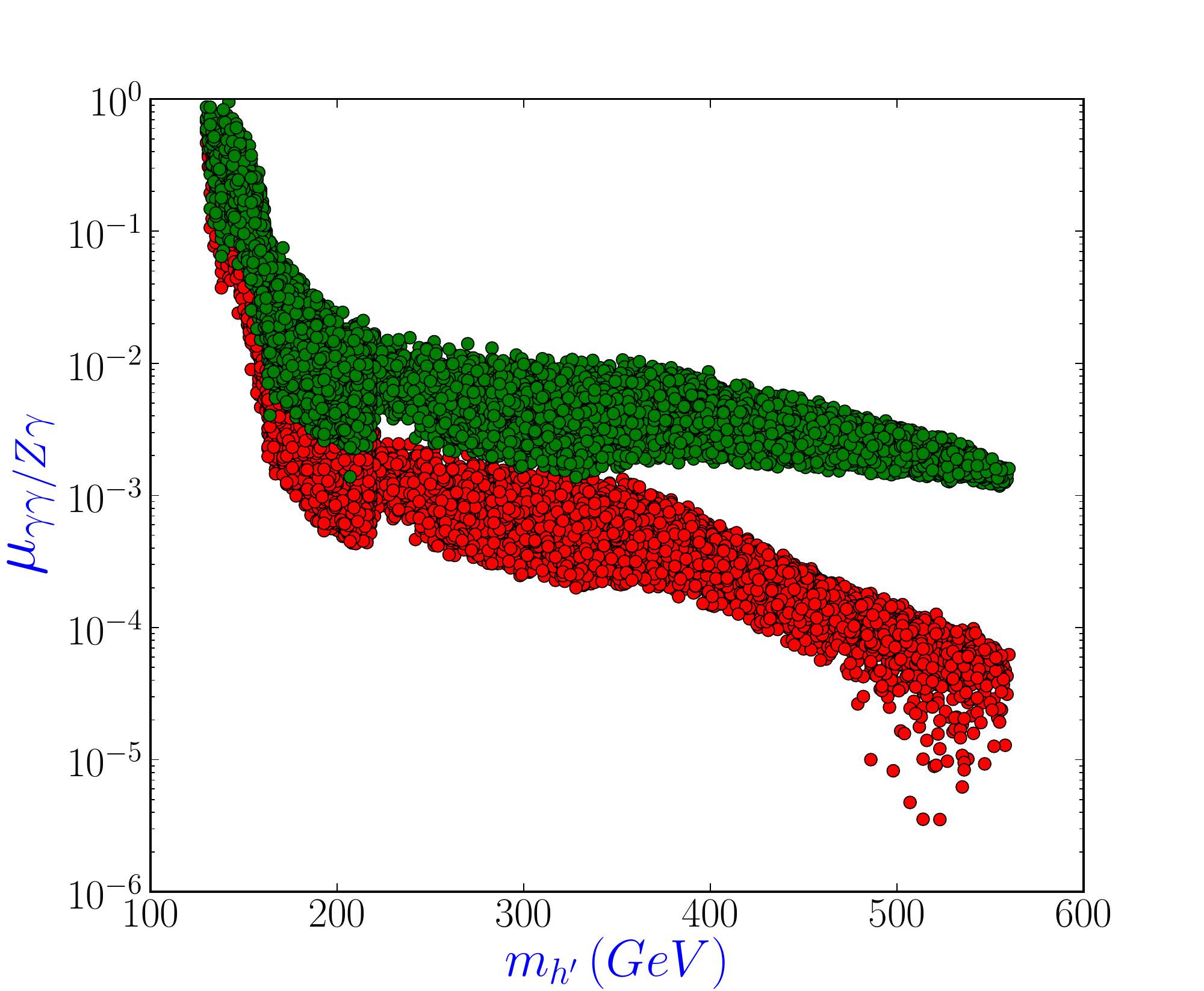}
\caption{Signal strength of $h'\to \gamma\gamma$  (red) and $Z\gamma$ (green) versus the
$h'$ mass for the
 light stau (top) and
light chargino (bottom) scenario.  }
\label{fig:3}
\end{figure}

We now turn to the CP-even Higgs  bosons of the BLSSM. Recall that the $h$ and $H$ states of the BLSSM are essentially
the same as in the MSSM. Furthermore, as shown in \cite{Abdallah:2014fra}, also the genuine BLSSM states, $h'$ and $H'$,  show a strong hierarchy, $m_{H'}\gg m_{h'}$, and the $h'$ can be the second lightest Higgs state, with mass just
larger than the SM-like state $h$. The enhancement of $ h \to \gamma\gamma$ with staus in SUSY models with extended gauge sector, first studied in \cite{Basso:2012tr}.
Fig.~\ref{fig:3} shows the signal strenghts (normalised to the SM as usual)
for the $h'\to \gamma\gamma$  and $Z\gamma$ modes versus
$m_{h'}$, again for light staus and charginos separately. In both cases the rates generally obtained are significantly higher than for
the case of the $H$ state (Fig. \ref{fig:2}), so as to favourably conclude that a $h'$ Higgs boson may well be within the reach of
the LHC run 2 for standard luminosities, also thanks to the rather light values that $m_{h'}$ can attain,
starting here as low as 135 GeV, thus also greatly enhancing its production rates with respect to the $H$ one (as $m_H \gsim
180$ GeV). We find that the $\gamma\gamma$ decay rates are larger than the $Z\gamma$ ones by
over an order of magnitude for light staus if $\tan\beta\sim 40$ whereas in the case of light charginos and low
$\tan\beta$ (5 and below) the hierarchy
between the two decay modes is inverted as the $Z\gamma$ one is largely dominant  over the $\gamma\gamma$ one
(even up to two orders of magnitude for heavy $h'$s).

{From Figs.~\ref{fig:2} and \ref{fig:3}, it is thus remarkable that for $H$ and $h'$ masses larger than 135 GeV the signal strength $\mu_{Z\gamma}$ can become larger than $\mu_{\gamma \gamma}$, unlike the expectation of the SM-like Higgs state $h$. This can be understood as follows. With heavy Higgs bosons, the $t$-loop function mediating the Higgs decay into $\gamma \gamma$ is increased while the $W$-loop function is decreased. Therefore, the net result for $\mu_{\gamma \gamma}$ is to be reduced significantly (up to three orders of magnitude) with respect to the $h$ case.
In contrast, the enhancement of the $t$-loop function and reduction of the $W$-loop one that mediate the Higgs decay to $Z \gamma$ are quite small, thus the corresponding values for $\mu_{Z \gamma}$ remain of the same order as those for $m_{h}=125$ GeV.}


In summary, we have shown that a comparative study of the $\gamma\gamma$ and $Z\gamma$ decay channels
of the SM-like Higgs boson discovered recently at the LHC may hold the key to unlock the door towards the understanding
of its nature, in the ultimate attempt to extract the underlying EWSB mechanism. If the latter is dynamically onset by
SUSY and no evidence of sparticle states exists from direct searches, an indirect proof of this paradigm may be obtained
by measuring the relative yield of Higgs event rates in the $\gamma\gamma$ and $Z\gamma$ decay modes.
On the one hand, a simultaneous enhancement of both with respect to the SM rates may be associated with the
presence of a light chargino. On the other hand,
the relative increase of the former with respect to the latter, with both being beyond the SM rates, may be induced
by a light stau. Under these circumstances, in the light of a degeneracy existing between the Higgs sectors of
the two SUSY realisations, such effects may equally be ascribed to either the MSSM or the BLSSM. What would
enable one to split the two SUSY scenarios would be the prompt detection within the BLSSM of a second Higgs signal in  $\gamma\gamma$ and/or $Z\gamma$ whereas this would not be possible in the MSSM.
Finally, the very distinctive hierarchy emerging in the $\gamma\gamma$ and $Z\gamma$ decay widths of the $h'$ state
may yield information about the structure of the BLSSM sparticle sector.

\vspace*{-0.75truecm}
\acknowledgments
\vspace*{-0.25truecm}
AH thanks W. Abdallah and M. Hemeda for fruitful discussions. The work of AH and SK is partially supported by the ICTP grant AC-80 while SM through the NExT Institute. The work of SK and SM is also funded through the
grant H2020-MSCA-RISE-2014 no. 645722 (NonMinimalHiggs).

\end{document}